\documentclass{skads2009}
\usepackage{graphicx}
\begin{document}
   \title{Wide Field Polarimetry and Cosmic Magnetism}

   \author{Rainer Beck\inst{1}
          }

   \institute{Max-Planck-Institut f\"ur Radioastronomie,
   Auf dem H\"ugel 69, 53121 Bonn, Germany
   \thanks{This work was supported by the
   European Commission Framework Program 6, Project SKADS, Square
   Kilometre Array Design Studies (SKADS), contract no 011938.} }

\abstract{The SKA and its precursors will open a new era in the
observation of cosmic magnetic fields and help to understand their
origin. In the SKADS polarization simulation project, maps of
polarized intensity and RM of the Milky Way, galaxies and halos of
galaxy clusters were constructed, and the possibilities to measure
the evolution of magnetic fields in these objects were investigated.
The SKA will map interstellar magnetic fields in nearby galaxies and
intracluster fields in nearby clusters in unprecedented detail.
All-sky surveys of Faraday rotation measures (RM) towards a dense
grid of polarized background sources with the SKA and ASKAP (POSSUM)
are dedicated to measure magnetic fields in distant intervening
galaxies, cluster halos and intergalactic filaments, and will be
used to model the overall structure and strength of the magnetic
fields in the Milky Way and beyond. Simple patterns of regular
fields in galaxies or cluster relics can be recognized to about
100~Mpc distance, ordered fields in unresolved galaxies or cluster
relics to redshifts of $z\simeq0.5$, turbulent fields in starburst
galaxies or cluster halos to $z\simeq3$, and regular fields in
intervening galaxies towards QSOs to $z\simeq5$. }
   \maketitle
%
%
\section{Origin of magnetic fields}

The origin of the first magnetic fields in the early Universe is
still a mystery (Widrow \cite{widrow02}). A large-scale primordial
field is hard to maintain in a young galaxy because the galaxy
rotates differentially, so that field lines get strongly wound up
during galaxy evolution, while observations show significant pitch
angles. ``Seed'' fields could also originate from the time of
cosmological structure formation by the Weibel instability in shocks
(Lazar et al. \cite{lazar09}) or could have been injected by the
first stars or jets generated by the first black holes (Rees
\cite{rees05}), followed by a mechanism to amplify and organize the
magnetic field.

The most promising mechanism to sustain magnetic fields in the
interstellar medium of galaxies is the dynamo (Beck et al.
\cite{beck96}). In young galaxies without ordered rotation a
small-scale dynamo (Brandenburg \& Subramanian \cite{brand05})
possibly amplified the seed fields from the protogalactic phase to
the energy density level of turbulence within less than $10^9$~yr.
To explain the generation of large-scale fields in galaxies, the
mean-field dynamo has been developed. It is based on turbulence,
differential rotation and helical gas flows ($\alpha$ effect),
generated by supernova explosions (Gressel et al. \cite{gressel08})
or by cosmic-ray driven Parker loops (Hanasz et al.
\cite{hanasz09}). The mean-field dynamo in galaxy disks predicts
that within a few $10^9$~yr large-scale regular fields are excited
from the seed fields (Arshakian et al. \cite{arshakian09}, see also
this volume), forming patterns (``modes'') with different azimuthal
symmetry in the disk and vertical symmetry in the halo.

The mean-field dynamo generates large-scale helicity with a non-zero
mean in each hemisphere. As total helicity is a conserved quantity,
the dynamo is quenched by the small-scale fields with opposite
helicity unless these are removed from the system (Shukurov et al.
\cite{shukurov06}). Outflows are essential for an effective
mean-field dynamo.

The magnetic fields in the intracluster medium could be seeded by
outflows from starburst galaxies (Donnert et al. \cite{donnert08})
or from AGNs and amplified by turbulent wakes, cluster mergers or a
turbulent dynamo (Subramanian et al. \cite{subra06}, Ryu et al.
\cite{ryu08}).

The fundamental questions are:
\begin{itemize}
\item When were the first magnetic fields generated: in young
galaxies, in protogalactic clouds, or are they relics from the early
Universe before the galaxies were formed?
\item How and how fast were magnetic fields amplified in the
interstellar and intracluster media?
\item Did magnetic fields affect the evolution of galaxies and
galaxy clusters?
\item How important are magnetic fields for the physics of galaxies,
like the efficiency to form stars from gas, the formation of spiral
arms or the generation of gas outflows?
\item How important are magnetic fields for the physics of the
intracluster medium, like cosmic-ray transport or anisotropic heat
conduction?
\item How strong and how ordered are magnetic fields in intergalactic
space?
\end{itemize}

\section{Measuring magnetic fields}

The intensity of synchrotron emission is a measure of the number
density of cosmic-ray electrons in the relevant energy range and of
the strength of the total magnetic field component in the sky plane.
Polarized emission emerges from ordered fields. As polarization
``vectors'' are ambiguous by $180\degr$, they cannot distinguish
{\em regular (coherent) fields} with a constant direction within the
telescope beam from {\em anisotropic fields}\ which are generated
from turbulent magnetic fields by compressing or shearing gas flows
and frequently reverse their direction on small scales. Unpolarized
synchrotron emission indicates {\em turbulent fields}\ with random
directions which have been tangled or generated by turbulent gas
flows. Only regular fields can give rise to Faraday rotation, while
anisotropic and random fields do not. Measurements of the Faraday
rotation from multi-wavelength observations allow to determine the
strength and direction of the regular field component along the line
of sight.

\begin{figure}
\centering
\includegraphics[bb = 26 28 378 373,width=5cm,clip=]{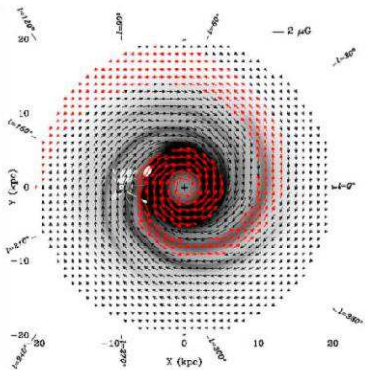}
\caption{The axisymmetric model of the large-scale structure of
magnetic fields in the Milky Way disk, derived from polarization
surveys and rotation measures of extragalactic sources (from Sun et
al. \cite{sun08}). } \label{fig:sun}
\end{figure}

A {\em grid of Faraday rotation measures (RM) measurements}\ towards
polarized background sources with small angular sizes is hardly
affected by Faraday depolarization within the foreground and a
powerful tool to study magnetic field patterns in galaxies (Stepanov
et al. \cite{stepanov08}) and in clusters (Krause et al.
\cite{krause+09}). A large number of background sources is required
to recognize the field patterns, to separate the Galactic foreground
contribution and to account for the intrinsic RM of extragalactic
sources.

The method of Fourier transform of multi-channel
spectro-polarimetric data into RM space by {\em RM Synthesis}\
(Brentjens \& de Bruyn \cite{brentjens05}) is going to revolutionize
radio polarization observations. It is able to separate RM
components from distinct foreground and background regions and hence
in principle to measure the 3-D structure of the magnetized
interstellar medium in galaxies. If the medium has a relatively
simple structure, e.g. a few emitting and Faraday-rotating regions,
{\em Faraday tomography}\ will become possible.
The distribution of the frequency channels across the total band and
the channel width of the observation defines the {\em Rotation
Measure Spread Function (RMSF)} (Heald \cite{heald09}). Cleaning of
the data cube with help of the known RMSF (``dirty beam'') is
similar to cleaning of synthesis data (Heald et al.
\cite{heald+09}).

\section{Magnetic fields in the Milky Way}

\begin{figure}
\begin{center}
\includegraphics[bb = 32 30 119 123,width=5cm,clip=]{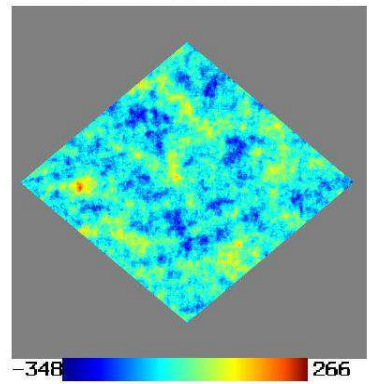}
\caption{Simulation of Faraday rotation measures at 1.6'' resolution
in a field of about $1.5\degr$ x $1.5\degr$ at l=$130\degr$,
b=$1\degr$. The mean RM is -86~rad m$^{-2}$, the RM dispersion
63~rad m$^{-2}$ and the structure function has a slope of about 0.7
(from Sun \& Reich \cite{sun09}). } \label{fig:RM_plane}
\end{center}
\end{figure}

\begin{figure}
\begin{center}
\includegraphics[bb = 32 30 119 123,width=5cm,clip=]{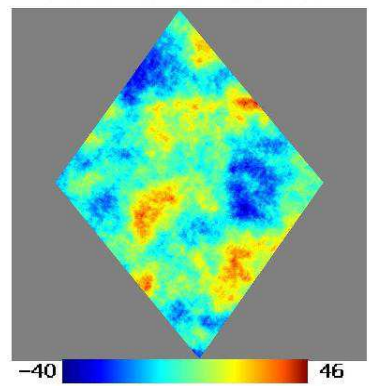}
\caption{Simulation of Faraday rotation measures at 1.6'' resolution
in a field of about $1.5\degr$ x $1.5\degr$ at l=$138\degr$,
b=$+70\degr$. The mean RM is +3~rad m$^{-2}$, the RM dispersion
13~rad m$^{-2}$ and the structure function has a slope of about 0.9
(from Sun \& Reich \cite{sun09}). } \label{fig:RM_high}
\end{center}
\end{figure}

Sun et al. (\cite{sun08}) used the all-sky maps of polarized
synchrotron emission at 1.4~GHz from the Milky Way (Wolleben et al.
\cite{wohl06}, Testori et al. \cite{testori08}) and at 22.8~GHz from
WMAP (Page et al. \cite{page07}) and the new Effelsberg Faraday RM
survey of polarized extragalactic sources (Han et al., in prep.)
were used to model the large-scale Galactic field, as part of the
SKADS science simulations (Fig.~\ref{fig:sun}). A large-scale
reversal is required about 1--2~kpc inside the solar radius, which
also agrees with the detailed study of RMs from extragalactic
sources near the Galactic plane (Brown et al. \cite{brown07}). More
large-scale reversals may exist (Han et al. \cite{han06}). The local
field is symmetric with respect to the Galactic plane, while the
toroidal component of the halo field is probably antisymmetric.
Models of a simple large-scale field structure of the Milky Way
(e.g. by Han et al. \cite{han02}) could not be confirmed by
statistical tests (Men et al. \cite{men08}). Similar to external
galaxies, the Milky Way's regular field probably has a complex
structure which can only be revealed by a larger sample of RM data
from pulsars and extragalactic sources. The RM sample derived from
the NVSS includes more than 37\,000 polarized sources, but it is
based on only two channels with a small frequency separation (Taylor
et al. \cite{taylor09}).


Detailed SKADS simulations of the diffuse Galactic emission towards
various Galactic directions were performed by Sun \& Reich
(\cite{sun09}). The turbulent field was assumed to be of Kolmogorov
type with an inner scale of 0.025~pc and and outer one of 10~pc. The
fields near the Galactic plane (Fig.~\ref{fig:RM_plane}) and at high
Galactic latitude (Fig.~\ref{fig:RM_high}) are different in mean RM,
RM dispersion and the slope of their structure functions. These
simulations will help to estimate the Galactic foreground
contribution and depolarization of extragalactic RM measurements
with the SKA, and are also useful for the foreground subtraction of
experiments to detect signals from the Epoch of Reionization.

\section{Magnetic fields in spiral galaxies}

\begin{figure}
\centering
\includegraphics[bb = 32 99 537 702,width=6cm,clip=]{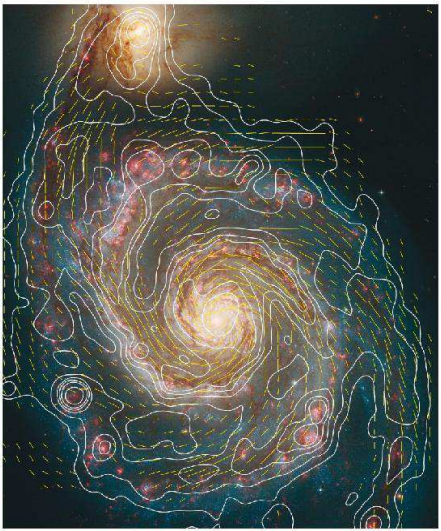}
\caption{Total radio emission (contours) and $B$--vectors of M~51,
combined from observations at 6~cm wavelength with the VLA and
Effelsberg telescopes and smoothed to 15'' resolution , overlaid
onto an optical image from the HST. Copyright: MPIfR Bonn and
\textit{Hubble Heritage Team}. Graphics: \textit{Sterne und
Weltraum} (from Fletcher et al. \cite{fletcher10}). }
\label{fig:m51}
\end{figure}

The ordered (regular and/or anisotropic) fields traced by the
polarized synchrotron emission are generally strongest
(10--15~$\mu$G) in the regions {\em between}\ the optical spiral
arms and oriented parallel to the adjacent spiral arms, in some
galaxies forming {\em magnetic arms}. These are probably generated
by a large-scale dynamo (Beck et al. \cite{beck96}). In galaxies
with strong density waves some of the ordered field is concentrated
on the inner edge of the spiral arms (Fig.~\ref{fig:m51}). The
ordered magnetic field forms spiral patterns in almost every galaxy
(Beck \cite{beck05}).

\begin{figure}
\centering
\includegraphics[bb = 104 142 464 645,width=4.5cm,clip=]{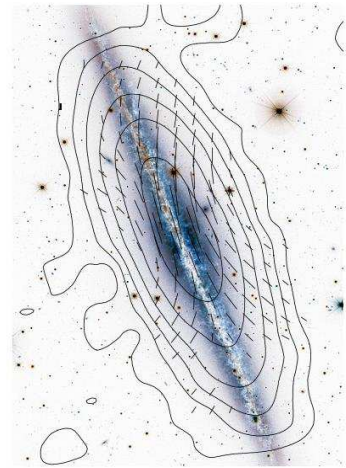}
\caption{Total radio emission (84'' resolution) and B--vectors of
the edge-on spiral galaxy NGC~891, observed at 3.6~cm wavelength
with the Effelsberg 100m telescope. The background optical image is
from the CFHT. Copyright: MPIfR Bonn and CFHT/Coelum (from Krause
\cite{krause09}). } \label{fig:n891}
\end{figure}

Spiral fields can be generated by compression at the inner edge of
spiral arms, by shear in interarm regions, or by dynamo action.
Large-scale patterns of Faraday rotation measures (RM) are
signatures of coherent dynamo fields and can be identified from
polarized emission of the galaxy disks (Krause \cite{krause90}) or
from RM data of polarized background sources (Stepanov et al.
\cite{stepanov08}). The Andromeda galaxy M~31 hosts a dominating
axisymmetric disk field (Fletcher et al. \cite{fletcher04}), as
predicted by dynamo models. Other candidates for a dominating
axisymmetric disk field are the nearby spiral IC~342 (Krause et al.
\cite{krause89}) and the irregular Large Magellanic Cloud (LMC)
(Gaensler et al. \cite{gaensler05}). However, in many observed
galaxy disks no clear patterns of Faraday rotation were found.
Either the field structure cannot be resolved with present-day
telescopes or the timescale for the generation of large-scale modes
is longer than the galaxy's lifetime (Arshakian et al.
\cite{arshakian09}).

Nearby galaxies seen edge-on generally show a disk-parallel field
near the disk plane. High-sensitivity observations of edge-on
galaxies like NGC~891 (Fig.~\ref{fig:n891}) and NGC~253 (Heesen et
al. \cite{heesen09}) revealed vertical field components in the halo
forming an ``X-shaped'' pattern. The field is probably transported
from the disk into the halo by an outflow emerging from the disk.

As another part of the SKADS science simulations, dynamo theory was
used to derive the timescales of amplification and ordering of
magnetic fields in galaxies. Based on models describing the
formation and evolution of dwarf and disk galaxies, an evolutionary
model of turbulent and regular magnetic fields was developed that
can be tested observationally (Arshakian et al. \cite{arshakian09}
and this volume):
\begin{itemize}
\item Strong turbulent fields (unpolarized synchrotron emission) can
be observed in galaxies at $z<10$.
\item Strong regular (coherent) fields (polarized synchrotron emission
and RM) can be observed in Milky Way-type galaxies at $z\le3$.
\item Large-scale patterns of fully coherent regular fields (polarized
synchrotron emission and large-scale RM patterns) can be observed in
dwarf and Milky-Way type galaxies at $z\le1$.
\item Giant galaxies (disk radius $>15$~kpc) may not have generated
fully coherent fields.
\item Major mergers enhance turbulent fields but destroy regular
fields and thus delay the formation of fully coherent fields.
\end{itemize}

\begin{figure}
\centering
\includegraphics[bb = 31 31 248 329,width=6cm,clip=]{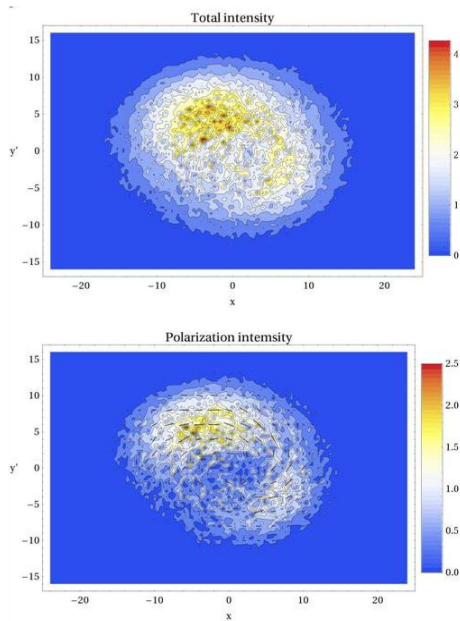}
\caption{Simulated map of the 5~GHz radio emission from a spiral
galaxy of 5~Gyr age. Top: total intensity, bottom: polarized
intensity (from Arshakian et al. in prep.). } \label{fig:galaxy}
\end{figure}

The derived timescales of amplification of amplitude and ordering of
the regular field and the sizes and star-formation rates of disk
galaxies from the SKADS galaxy simulations (Obreschkow et al.
\cite{obreschkow09}) were used to simulate the evolution of total
radio intensity, polarization and Faraday rotation with age of a
galaxy ($z\le 3$) at frequencies from 150~MHz to 18~GHz
(Fig.~\ref{fig:galaxy}).

About 10 RM values in the area of a foreground galaxy are already
sufficient to recognize a simple large-scale field structure (if the
Galactic foreground can be properly subtracted), while more than
1000 values are required for a 3-D field reconstruction (Stepanov et
al. \cite{stepanov08}).

Ordered fields of nearby galaxies seen edge-on near the disk plane
are preferably oriented parallel to the plane (Krause
\cite{krause09}). Polarized emission can be detected from unresolved
galaxies if the inclination is larger than about $20\degr$ (Stil et
al. \cite{stil09}). This opens a new method to search for ordered
fields in distant galaxies.

\section{Galaxy clusters}

\begin{figure}
\centering
\includegraphics[bb = 26 27 279 338,width=5cm,clip=]{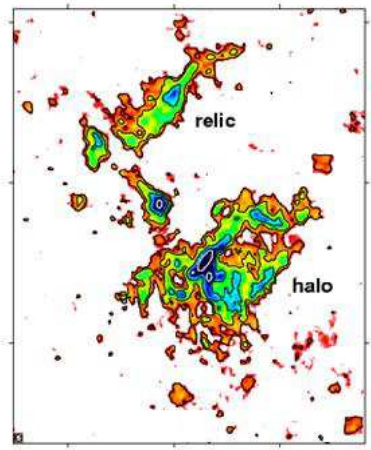}
\caption{Radio emission at 1.4~GHz from the cluster A~2255. The
field size is about 20' x 25' (from Govoni et al. \cite{govoni05}).
} \label{fig:a2255}
\end{figure}

Some fraction of galaxy clusters, mostly the X-ray bright ones, has
diffuse radio emission (Cassano et al. \cite{cassano08}), emerging
from diffuse {\em halos}\ and steep-spectrum {\em relics}\
(Fig.~\ref{fig:a2255}). Radio halos are mostly unpolarized and
emerge from turbulent intracluster magnetic fields, observed as
dispersion in RM decreasing with distance from the cluster center
(Clarke et al. \cite{clarke01}). Relics can emit highly polarized
radio waves from anisotropic magnetic fields generated by
compression in merger shocks (En{\ss}lin et al. \cite{ensslin98}). A
polarized region of about 1~Mpc size was discovered in Abell~2255
(Govoni et al. \cite{govoni05}).

Equipartition strengths of the total magnetic field range from 0.1
to $1~\mu$G in halos and are higher in relics. On the other hand,
Faraday rotation data towards background sources behind cluster
halos reveals fields of a few $\mu$G strength fluctuating on
coherence scales of a few kpc (Govoni \& Feretti \cite{govoni04})
and even $40~\mu$G in the cores of cooling flow clusters (Carilli \&
Taylor \cite{carilli02}) where they may be dynamically important.
The reason for the discrepancy in field strengths is still under
discussion.

High-resolution RM maps of radio galaxies embedded in a cluster
allowed to derive the power spectra of the turbulent intracluster
magnetic fields which are of Kolmogorov type and have coherence
scales of about 1--5~kpc (Vogt \& En{\ss}lin \cite{vogt03}).

\begin{figure}
\centering
\includegraphics[bb = 69 42 385 280,width=7cm,clip=]{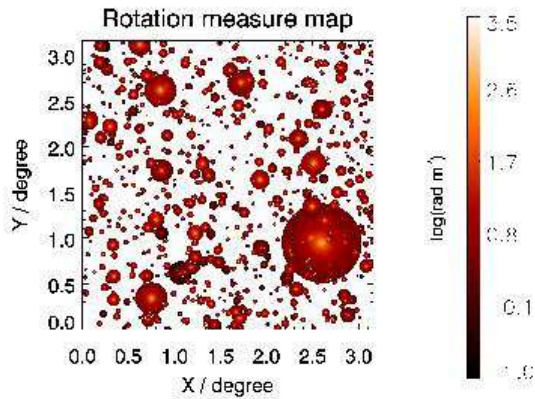}
\caption{Simulation of RMs from a cosmological cluster model (from
Krause et al. \cite{krause+09}). } \label{fig:cluster}
\end{figure}

\begin{figure}
\centering
\includegraphics[bb = 38 16 572 685,width=5cm,angle=270,clip=]{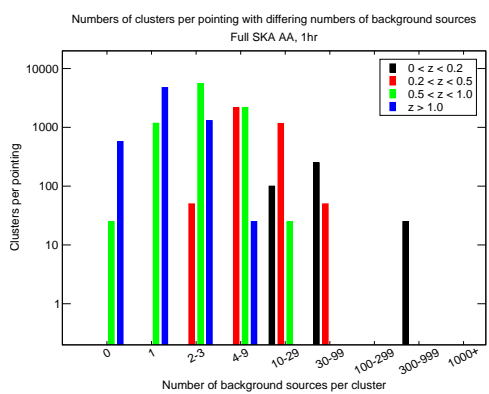}
\caption{Simulation of the number of RMs per cluster observable with
the SKA Aperture Array (300--1000~MHz) per field of view
(250~deg$^2$) within 1~h integration time (from Krause et al.
\cite{krause+09}). } \label{fig:cluster_SKA}
\end{figure}

Within the SKADS simulation project, Krause et al. (\cite{krause09})
derived RM maps expected for a cluster population from an analytical
cosmological model (Fig.~\ref{fig:cluster}). About 12\% of the sky
is covered with regions of $|RM|>$~10~rad m$^{-2}$, and 66\% of all
clusters are located at $z\le1$. From this model and a model for the
number density of polarized background sources, the number of
background sources per cluster to be observed with the SKA Aperture
Array was computed (Fig.~\ref{fig:cluster_SKA}). The number of RMs
for clusters out to z=1 will be sufficiently high to map the
magnetic field distribution. At larger redshifts deeper integration
will be needed. The total number of detectable clusters is about
$10^4$ per field of view and per hour of integration. Measurement of
the decrease of the maximum cluster RM with redshift can be used to
constrain the magnetic field evolution $B\propto(1+z)^n$ within an
accuracy of $\Delta n< 0.4$.

\section{Intergalactic filaments}

The prediction of a large-scale ``Cosmic Web'' is one of the
defining characteristics of large-scale structure simulations.
Moreover, galaxies and the intra-cluster medium account for only
approximately one third of the baryon density in the local Universe
expected in a concordance cosmology. The majority of the missing
baryons are likely to reside in a warm-hot intergalactic medium
(WHIM) which in turn is expected to reside in the cosmic web of the
large-scale structure. Galaxy clusters, the brightest knots of the
Cosmic Web, are filled with relativistic particles and magnetic
fields, as observed through the radio synchrotron emission.

The search for magnetic fields in the intergalactic medium (IGM) is
of fundamental importance for cosmology. All ``empty'' space in the
Universe may be magnetized. Its role as the likely seed field for
galaxies and clusters and its possible relation to structure
formation in the early Universe, places considerable importance on
its discovery. Models of structure formation predict strong
intergalactic shocks which enhance the field.

Various mechanisms have been suggested for the origin of a magnetic
field in the cosmic web. The field could be produced via the Weibel
instability (a small-scale plasma instability) at structure
formation shocks (Medvedev et al. \cite{medvedev04}). Another
possibility is that the field is injected into the WHIM via the
action of injection from galactic black holes (AGNs) and other
outflows. In each case the field is subsequently amplified by
compression and large-scale shear-flows (Br\"uggen et al.
\cite{brueggen05}). Ryu et al. (\cite{ryu08}) have argued that
highly efficient amplification is possible via MHD turbulence, with
the source of the turbulent energy being the structure formation
shocks themselves (Fig.~\ref{fig:ryu}). Estimates of the strength of
the turbulent field in filaments obtained from MHD simulations with
a primordial seed field range typically between 0.1~$\mu$G and
0.01~$\mu$G, while regular fields are weaker.

To date there has been no detection of a general magnetic field in
the IGM. In an intergalactic region of about $2\degr$ extent west of
the Coma Cluster, containing a group of radio galaxies, enhanced
synchrotron emission yields an equipartition total field strength of
$0.2-0.4~\mu$G (Kronberg et al. \cite{kronberg07}). Xu et al.
(\cite{xu06}) observed an excess of rotation measures (RM) towards
two super-clusters which may indicate regular magnetic fields of
$<0.3~\mu$G on scales of order 500~kpc. Lee et al. (\cite{lee09})
found a statistical correlation at the 4$\sigma$ level of the RMs of
background sources with the galaxy density field and derived a 30~nG
intergalactic field with about 1~Mpc coherence length.

\section{Prospects}

Future radio telescopes will widen the range of observable magnetic
phenomena. High-resolution, deep observations at high frequencies,
where Faraday effects are small, require a major increase in
sensitivity for continuum observations which will be achieved by the
Extended Very Large Array (EVLA) and the planned Square Kilometre
Array (SKA). The detailed structure of the magnetic fields in the
ISM of galaxies, in galaxy halos, in cluster halos and in cluster
relics can then be observed. The turbulence power spectra of the
magnetic fields can be measured (Vogt \& En{\ss}lin \cite{vogt03}).
Direct insight into the interaction between gas and magnetic fields
in these objects will become possible. The SKA will also allow to
measure the Zeeman effect in much weaker magnetic fields in the
Milky Way and in nearby galaxies.

\begin{figure}
\centering
\includegraphics[bb = 0 0 566 425,width=9cm,clip=]{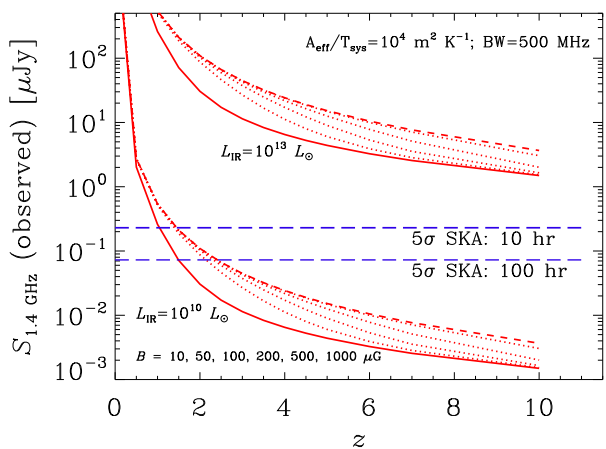}
\caption{Synchrotron emission at 1.4~GHz as a function of redshift
$z$ and magnetic field strength $B$, and the $5\sigma$ detection
limits for 10~h and 100~h with the SKA (from Murphy
\cite{murphy09}). } \label{fig:murphy}
\end{figure}

Detection of polarized emission from distant, unresolved galaxies
indicates large-scale ordered fields (Stil et al. \cite{stil09}),
and statistics can be compared with the predictions of dynamo theory
(Arshakian et al. \cite{arshakian09}). The SKA will detect Milky-Way
type galaxies at $z\le1.5$ (Fig.~\ref{fig:murphy}) and their
polarized emission at $z\le0.5$ (assuming 10\% percentage
polarization). Bright starburst galaxies can be observed at larger
redshifts, but are not expected to host ordered or regular fields.
Cluster relics are also detectable at large redshifts through their
integrated polarized emission. This effect still has to be
investigated.

Unpolarized synchrotron emission, signature of turbulent magnetic
fields, can be detected with the SKA out to very large redshifts for
starburst galaxies, depending on luminosity and magnetic field
strength (Fig.~\ref{fig:murphy}), and also for cluster halos.
However, for fields weaker than $3.25~\mu$G $(1+z)^2$, energy loss
of cosmic-ray electrons is dominated by the inverse Compton effect
with CMB photons, so that their energy appears mostly in X-rays and
not in the radio range. On the other hand, for strong fields the
energy range of the electrons emitting at a 1.4~GHz drops to low
energies, where ionization and bremsstrahlung losses may become
dominant. In summary, the mere detection of synchrotron emission at
high redshifts will constrain the range of allowed magnetic field
strengths.

Forthcoming low-frequency radio telescopes like the Low Frequency
Array (LOFAR), Murchison Widefield Array (MWA), Long Wavelength
Array (LWA) and the low-frequency SKA will be suitable instruments
to search for extended synchrotron radiation at the lowest possible
levels in outer galaxy disks and the transition to intergalactic
space (Beck \cite{beck09}) and in steep-spectrum cluster halos
(Brunetti et al. \cite{brunetti08}).

If polarized emission is too weak to be detected, the method of {\em
RM grids}\ towards background QSOs can still be applied. Here, the
distance limit is given by the polarized flux of the background QSO
which can be much higher than that of the intervening galaxy. A
reliable model for the global structure of the magnetic field of
nearby galaxies needs a large number of RM values from a large
number density of polarized background sources, hence large
sensitivity and/or high survey speed. The ``POSSUM-Wide'' survey at
1.4~GHz with the planned Australia SKA Pathfinder (ASKAP) telescope
with 30~deg$^2$ field of view (Gaensler et al. \cite{gaensler10})
will measure about 100 RM values from polarized extragalactic
sources per square degree within 10~h integration time, assuming a
slope of the cumulative source count function of $\gamma=-0.9$
(Stepanov et al. \cite{stepanov08}). Similarly deep integrations
with the EVLA and with MeerKAT will show about 5~times more sources,
but their fields of view are small and will allow only a limited
number of pointings.

\begin{figure}
\centering
\includegraphics[bb = 19 27 336 331,width=6cm,clip=]{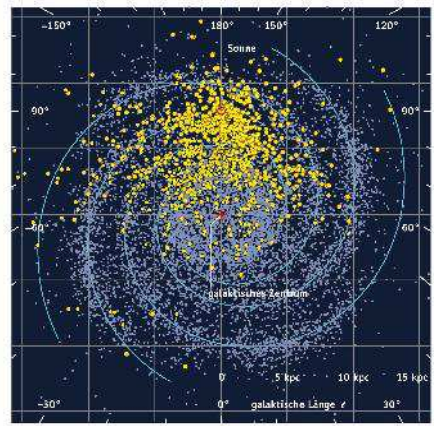}
\caption{Simulation of about 20\,000 new pulsars (blue) in the Milky
Way that will be detected with the SKA, compared to about 2000
pulsars known today (yellow). Graphics: \textit{Sterne und Weltraum}
(from Cordes, priv. comm). } \label{fig:pulsars}
\end{figure}

The SKA pulsar survey will find about 20\,000 new pulsars which will
be mostly polarized and reveal RMs (Fig.~\ref{fig:pulsars}),
perfectly suited to measure the Milky Way's magnetic field with
extremely high precision.

\begin{figure}
\centering
\includegraphics[bb = 147 78 450 717,width=3.5cm,angle=270,clip=]{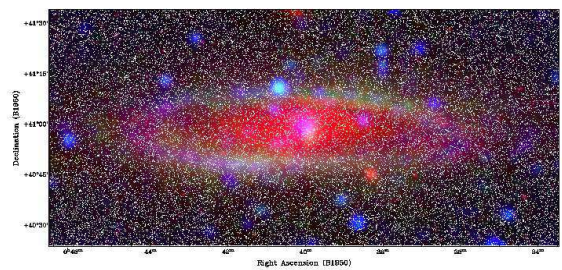}
\caption{Simulation of the number of RMs (about 10\,000) in the
region of M~31 observable with the SKA within 1~h integration time
(from Gaensler, priv. comm.). } \label{fig:m31_sim}
\end{figure}

The SKA ``Magnetism'' Key Science Project plans to observe a
wide-field survey (at least $10^4$~deg$^2$) around 1~GHz with 1~h
integration per field which will be able to detect sources with
0.5--1~$\mu$Jy flux density and measure at least 1500 RMs per square
degree. This will contain at least $2\times10^7$ RMs from compact
polarized extragalactic sources at a mean spacing of $\simeq90''$,
plus at least $10^4$ RM values from pulsars with a mean spacing of
$\simeq30'$ (Gaensler et al. \cite{gaensler04}, \cite{gaensler09}).
If the cumulative source count function of polarized sources has a
slope as steep as $\gamma=-1.1$ (Stepanov et al. \cite{stepanov08}),
the total RM number will be $10\times$ larger and the mean spacing
as low as $\simeq30''$. More than 10\,000 RM values are expected in
the area of M~31 (Fig.~\ref{fig:m31_sim}) and will allow the
detailed reconstruction of the 3-D field structure in this and many
other nearby galaxies, while simple patterns of regular fields can
be recognized out to distances of about 100~Mpc (Stepanov et al.
\cite{stepanov08}) where the polarized flux is too low to be mapped.
The magnetism of cluster halos can be measured by the RM grid to
redshifts of about 1 (Fig.~\ref{fig:cluster_SKA}).

The SKA ``Magnetism'' Key Science Project also plans a deep-field
survey (at least 40~deg$^2$) with 100~h integration per field (Stil
et al. \cite{stil+09}), to detect polarized sources with 50~nJy flux
density and measure their RMs. With at least $3\times10^4$ RMs per
square degree and at a mean spacing of 10--20'', about 3~times
larger distances can be reached than with the wide-field survey.

\begin{figure}
\centering
\includegraphics[bb = 18 144 592 718,width=6cm]{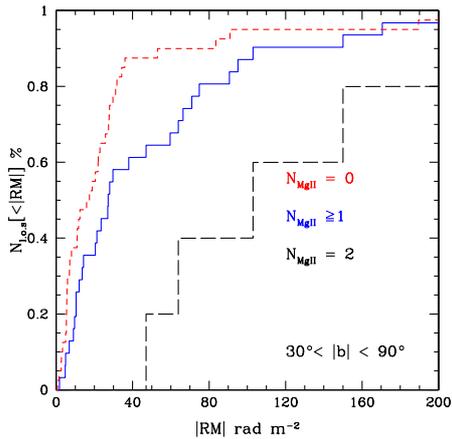}
\caption{Cumulative RM distributions for 31 sightlines towards QSOs
with optical spectra showing one or two strong MgII absorption line
systems (thick blue line), for 5 sightlines with two strong MgII
systems (black dashed line) and for 40 sightlines without strong
MgII absorption lines (thin red dashed line). The redshifts of the
intervening galaxies are between z=0.6 and 2 (from Bernet et al.
\cite{bernet08}). } \label{fig:bernet}
\end{figure}

Faraday rotation in the direction of QSOs allows to determine the
strength and pattern of a regular field in an intervening galaxy
(Kronberg et al. \cite{kronberg92}). Significant regular fields of
several $\mu$G strength were detected in distant galaxies
(Fig.~\ref{fig:bernet}). This method can be applied to distances of
young QSOs ($z\simeq5$), and the SKA will provide a large data
sample for excellent statistics. Mean-field dynamo theory predicts
RMs from regular galactic fields at $z\le3$ (Arshakian et al.
\cite{arshakian09}), but the RM values are reduced by the redshift
dilution factor of $(1+z)^{-2}$. If an overall IGM field with a
coherence length of a few Mpc existed in the early Universe and its
strength varied proportional to $(1+z)^2$ (Widrow \cite{widrow02}),
its signature may become evident at redshifts of $z>3$. Averaging
over a large number of RMs is required to see the IGM signal. Our
goal is to detect an IGM magnetic field of 0.1~nG at 5$\sigma$ in
the presence of Galactic foregrounds. Kolatt (\cite{kolatt98}) has
calculated that an RM density of $\approx1000$ sources per
deg$^{-2}$ is sufficient to meet this requirement.

\begin{figure}
\centering
\includegraphics[bb = 19 27 493 274,width=8cm,clip=]{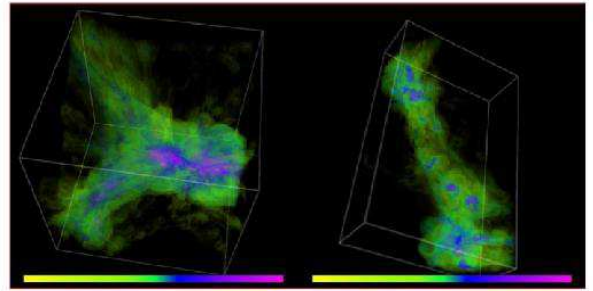}
\caption{Simulation of magnetic fields in the Cosmic Web (at $z =
0$) in a volume of $25\, h^{-1}$~Mpc$^3$ centered around a cluster
complex (left panel), and in a volume of $25 \times 15.6 \times
6.25\,h^{-1}$~Mpc$^3$ including a number of galaxy groups along a
filament (right panel). The color codes the magnetic field strength
(logarithmically scaled) from 0.1~nG (yellow) to 10~$\mu$G
(magenta). Clusters and groups are shown with magenta and blue,
while filaments are green (from Ryu et al. \cite{ryu08}). }
\label{fig:ryu}
\end{figure}

If the filaments of the local Cosmic Web outside clusters contain a
magnetic field (Fig.~\ref{fig:ryu}), possibly enhanced by IGM
shocks, we can hope to detect this field by direct observation of
its total synchrotron emission (Keshet et al. \cite{keshet04}) and
possibly its polarization, or by Faraday rotation towards background
sources. For fields of $\approx 10^{-8}-10^{-7}$~G with 1~Mpc
coherence length and $n_e\approx 10^{-5}$~cm$^{-3}$ electron
density, Faraday rotation measures between 0.1 and 1~rad m$^{-2}$
are expected. A 30~nG regular field on a coherence scale of 1~Mpc
(Lee et al. \cite{lee09}) would generate about 0.2~rad m$^{-2}$,
which probably cannot be detected directly. Promising is a
statistical analysis like the measurement of the power spectrum of
the magnetic field of the Cosmic Web (Kolatt \cite{kolatt98}) or the
cross-correlation with other large-scale structure indicators like
the galaxy density field. Detection of a general IGM field, or
placing stringent upper limits on it, will provide powerful
observational constraints on the origin of cosmic magnetism.

The SKA wide-field RM survey is planned in the frequency band of the
SKA Aperture Array (300--1000~MHz). At lower frequencies synchrotron
emission is stronger but depolarization effects are more severe.
With a model of depolarization by internal Faraday dispersion, the
wavelength of maximum polarized emission can be computed as a
function of RM dispersion (Arshakian et al., in prep.). Polarized
emission from galaxy disks should be observed around 1~GHz, while
galaxy halos and intracluster media can be best observed between
300~MHz and 1~GHz. A model of external depolarization of the
emission from background sources in a foreground cluster leads to a
similar result (Krause et al. \cite{krause+07}).




\section{Summary}

The SKA and its precursors will measure the structure and strength
of the magnetic fields in the Milky Way, in intervening galaxies and
clusters, and in the intergalactic medium. Looking back into time,
the future telescopes can shed light on the origin and evolution of
cosmic magnetic fields.

The observational methods are:

\begin{itemize}
\item 3-D RM grid from extragalactic sources and pulsars to map the
detailed 3-D structure of the Milky Way's magnetic field
\item High-resolution mapping of total and polarized synchrotron emission
from nearby galaxies, clusters halos and relics
\item Reconstruction of 3-D field patterns in nearby galaxies and nearby
clusters from RMs towards polarized background sources
\item Recognition of simple patterns of regular fields in galaxies from
RMs towards polarized background sources (at $z\le0.02$)
\item Detection of polarized synchrotron emission from distant galaxies
and distant halo relics (at $z\le0.5$)
\item Recognition of structure of turbulent fields in clusters from RMs
towards polarized background sources (at $z\le0.5$)
\item Detection of total synchrotron emission from distant starburst galaxies
and distant clusters (at $z\le3$)
\item Detection of regular fields in very distant intervening galaxies towards
QSOs (at $z\le5$).
\end{itemize}


\end{document}